\documentclass[prl,onecolumn,showpacs,amsmath,amssymb]{revtex4}
\usepackage{graphics}

\def\nuc#1#2{\relax\ifmmode{}^{#1}{\protect\text{#2}}\else${}^{#1}$#2\fi}

\begin{document}

\title{Measurement of two-halo neutron transfer reaction p(\nuc{11}{Li},\nuc{9}{Li})t at 3$A$ MeV}
\author{I. Tanihata}
\altaffiliation[present address: ]{RCNP, Osaka University, Mihogaoka, Ibaraki 567-0047, Japan.}
\author{M. Alcorta}
\altaffiliation[present address: ]{Institute de Estructura de la Materia, CSIC, Serrano 113bis, E-28006 Madrid, Spain.}
\author{D. Bandyopadhyay}
\author{R. Bieri}
\author{L. Buchmann}
\author{B. Davids}
\author{N. Galinski}
\author{D. Howell}
\author{W. Mills}
\author{R. Openshaw}
\author{E. Padilla-Rodal}
\author{G. Ruprecht}
\author{G. Sheffer}
\author{A. C. Shotter}
\author{S. Mythili}
\author{M. Trinczek}
\author{P. Walden}
\affiliation{TRIUMF, 4004 Wesbrook Mall, Vancouver, BC, V6T 2A3, Canada}

\author{H. Savajols}
\author{T. Roger}
\author{M. Caamano}
\author{W. Mittig}
\altaffiliation[present address: ]{NSCL, MSU East Lansing, MI 48824-1321, USA.}
\author{P. Roussel-Chomaz}
\affiliation{GANIL, Bd Henri Becquerel, BP 55027, 14076 Caen Cedex 05, France}

\author{R. Kanungo}
\author{A. Gallant}
\affiliation{Saint Mary's University, 923 Robie St., Halifax, Nova Scotia B3H 3C3, Canada}

\author{M.Notani}
\author{G. Savard}
\affiliation{ANL, 9700 S. Cass Ave., Argonne, IL 60439, USA}

\author{I. J. Thompson}
\affiliation{    LLNL, L-414, P.O. Box 808, Livermore CA 94551, USA}

\date{(received on  )}

\begin{abstract}
The p(\nuc{11}{Li},\nuc{9}{Li})t reaction has been studied for
the first time at an incident energy of 3$A$ MeV delivered by the new
ISAC-2 facility at TRIUMF. An active target detector MAYA, build at
GANIL, was used for the measurement. The differential cross sections
have been determined for transitions to the \nuc{9}{Li} ground and
the first excited states in a wide range of scattering angles.
Multistep transfer calculations using different \nuc{11}{Li} model
wave functions, shows that wave functions with strong correlations
between the halo neutrons are the most successful in reproducing the
observation.
\end{abstract}


\maketitle

The neutron-rich Li isotope \nuc{11}{Li} has the most pronounced
two-neutron halo. Presently the most important question about the
halo structure concerns the nature of the interaction and
correlation between the two halo neutrons. In a halo, the
correlation may be different from that of a pair of neutrons in
normal nuclei for several reasons. Halo neutrons are very weakly
bound and, therefore, the effect of the continuum becomes important.
The wave function of the halo neutrons has an extremely small
overlap with that of the protons and, thus, may experience
interactions much different from those of neutrons in normal nuclei.
The density of halo neutrons is very low compared with normal
nuclear density and, thus, may give rise to quite different
correlations from that in stable or near-stable nuclei. So far,
there have been several experimental attempts to elucidate the
nature of these correlations between the halo neutrons in
\nuc{11}{Li}. For example, measurements of neutrons and \nuc{9}{Li}
from the fragmentation of \nuc{11}{Li} have been used to determine
the momentum correlation between two halo neutrons \cite{ref-01}.
However, the contribution of the \nuc{10}{Li} resonance, which
decays to \nuc{9}{Li}+n immediately, made it difficult to reach
definitive conclusions. Later, Zinser et al. \cite{ref-02} studied
high-energy stripping reactions of \nuc{11}{Li} and \nuc{11}{Be} to
\nuc{10}{Li}, and the analyses of the momentum distributions
suggests the necessity of considerable mixing of $(1\text{s}_{1/2})^2$ and
$(0\text{p}_{1/2})^2$ configurations in the ground state of \nuc{11}{Li}.
The importance of the s-wave contribution is also seen in Coulomb
dissociation measurements \cite{ref-03,ref-04}. Determinations of
such amplitudes have also been attempted from data associated with
the beta-decay of \nuc{11}{Li}; however, no definite conclusions
could be reached.

A recent measurement of the charge radius of \nuc{11}{Li} and
\nuc{9}{Li} \cite{ref-05}, when combined with the matter radii of
\nuc{11}{Li} and \nuc{9}{Li} \cite{ref-06}, provides unique
information concerning the two-neutron distribution. The
root-mean-square(rms) distance between two halo neutrons
$\langle{}r^2_\text{{n-n}}\rangle^{1/2}$, as well as the distance between
the two-neutron center-of-mass and the center of the core
$\langle{}r^2_\text{{c-2n}}\rangle^{1/2}$, can be evaluated under two
assumptions: a) that \nuc{11}{Li} consists of a \nuc{9}{Li} core
plus two halo neutrons and b) that there is no angular correlation
between the position vector connecting the two-halo neutrons and the
position vector connecting the center of mass of the two-neutrons to
the center of the core. Taking these assumptions into account, and
using experimental data from references \cite{ref-05,ref-06}, we
obtain $\langle{}r^2_\text{{n-n}}\rangle^{1/2}=7.52\pm1.72$ fm and
$\langle{}r^2_\text{{c-2n}}\rangle^{1/2}=6.15\pm0.52$ fm. This shows that
the average opening angle of the two-halo neutrons, relative to the
core centre, is about 60 degrees. Similar analysis can be made for
\nuc{6}{He}, if again it is assumed:
$\nuc{6}{He}\rightarrow\nuc{4}{He}+\text{n}+\text{n}$. From this, and using
measured matter and charge radii of \nuc{4,6}{He}
\cite{ref-06,ref-vii}, we obtain
$\langle{}r^2_\text{{n-n}}\rangle^{1/2}=3.91\pm0.28$ fm, and
$\langle{}r^2_\text{{c-2n}}\rangle^{1/2}=3.84\pm0.06$ fm. The average
opening angle in this case is 54 degrees.

In \nuc{6}{He} case the
wave function of the halo is known better than that of \nuc{11}{Li}
because it is mainly of a $\text{p}_{3/2}$ character. From three-body
calculations two separate components (di-neutron and cigar shape)
contribute almost equally to the halo structure \cite{ref-viii}.
Experimental observations for the two-neutron transfer reaction,
$\nuc{6}{He}+\nuc{4}{He}\rightarrow\nuc{4}{He}+\nuc{6}{He}$,
(backward elastic scattering included) yields results consistent
with such theoretical calculations \cite{ref-09}. Although it seems
that the relation between $\langle{}r^2_\text{{n-n}}\rangle^{1/2}$ and
$\langle{}r^2_\text{{c-2n}}\rangle^{1/2}$ is similar between \nuc{11}{Li}
and \nuc{6}{He}, differences in the structure of the wave functions
are expected because of the complexity of the \nuc{11}{Li} wave
function, namely a mixing of two different orbitals ($1\text{s}_{1/2}$ and
$0\text{p}_{1/2}$). The mixing amplitude of these waves is an important
factor that determines the binding and the halo structure in
\nuc{11}{Li}.

The newly constructed ISAC-2 accelerator at TRIUMF now provides the
highest intensity beam of low-energy \nuc{11}{Li} up to 55 MeV. This
beam enabled the measurement of the two-neutron transfer reaction of
\nuc{11}{Li} for the first time. The reaction Q-value of
\nuc{11}{Li}(p,t)\nuc{9}{Li} is very large (8.2 MeV) and, thus,
the reaction channel is open at such low energies. The beam energy
used in this experiment (33 MeV) is not as high as usually used in
studies of direct reactions, nevertheless due to the low separation
energy of the two halo neutrons ($\sim$400 keV compared with about
10 MeV in stable nuclei) and low Coulomb barrier ($\sim$0.5 MeV) the
reaction is expected to be mainly direct. Momentum matching is also
good at this low energy because of the small internal momentum of
the halo neutrons.

The beam of \nuc{11}{Li} was accelerated to energy of 36.9 MeV. Beam
intensity on the target was about 2500 pps on average, and about
5000 pps at maximum. Measurement of the transfer reaction was made
possible at this low beam intensity through the use of the MAYA
active target detector brought to TRIUMF from GANIL. MAYA has a
target-gas detection volume (28 cm long in the beam direction, 25 cm
wide, and 20 cm high) for three-dimensional tracking of charged
particles, and a detector telescope array at the end of the chamber.
Each detector telescope consisted of a 700 $\mu$m thick Si detector
and a 1 cm thick CsI scintillation counter of $5\times5$ cm$^2$. The
array consists of twenty sets of telescopes. MAYA was operated with
isobutane gas first at a gas pressure of 137.4 mbar and then at 91.6
mbar. These two different pressure settings were used to cross check
the validity of the analysis by changing the drift speed of ionized
electrons and by changing the energy loss density.  The coverage of
center of mass angles was also different under these pressures - as
will be discussed later. Reaction events were identified by a
coincidence between a parallel plate avalanche chamber (PPAC), which
is placed just upstream of MAYA, and the Si array. \nuc{11}{Li} ions
that did not undergo a reaction were stopped in the blocking
material just before the Si array. Details of MAYA can be found in
Ref.\cite{ref-10}.

The two-neutron transfer reaction
p(\nuc{11}{Li},\nuc{9}{Li})t was identified by two methods
depending on the scattering angles. For forward scattering in the
c.m., \nuc{9}{Li} ions in the laboratory frame are emitted at small
angles and have sufficient energies to traverse the gas and hit the
Si array, and so \nuc{9}{Li} ions were identified by the
$\Delta{}E-E$ method. The $\Delta{}E$ signal was obtained from the
last 5 cm of the MAYA gas detector. Tritons emitted near 90 degrees
have low energies so that they stop within the gas detector and thus
provide total energy signals. However tritons emitted in smaller
angles, but larger than the angle covered by the Si array, punch
though the gas volume and therefore only scattering angles and
partial energy losses in the chamber could be measured. The major
background for such events came from the
$\nuc{11}{Li}+\text{p}\rightarrow\nuc{10}{Li}+\text{d}\rightarrow\nuc{9}{Li}+\text{n}+\text{d}$
reaction. Fortunately, the kinematical locus in a two-dimensional
plot between emission angles of \nuc{9}{Li} and light particles,
$[\theta(\text{Li})-\theta(\text{light})]$, of the (p,d) reaction
is well separated from the punch through (p,t) events.

For large angle scattering in the c.m., tritons were detected by the
Si and CsI detectors and identified clearly by the $\Delta{}E-E$
technique. Under this condition, \nuc{9}{Li} stops inside the gas
detector volume and thus the total energy, the range, and the
scattering angle were determined. Lithium ions can easily be
identified from their energy loss along the track, but
identification of the isotope mass number is more difficult. The
largest background events are due to scattered \nuc{11}{Li} with
accidental coincidence that have higher energies and will hit the Si
array. So rejection of such \nuc{11}{Li} events was easily
accomplished.  Use again was made of a
$\theta(\text{Li})-\theta(\text{light})$ correlation plot for final
identification of the (p,t) reaction. To remove other sources of
background, additional selections were applied. For forward angle
events, important correlations for these selections were:
$E(\text{Li})-\theta(\text{Li})$, $E(\text{Li})-\theta(\text{H})$,
and $Q_c(\text{H})-\theta(\text{H})$, where $E$ is the energy of a
particle determined by Si array and $Q_c$ is the total charge
collected in the gas detection volume. The last correlation was
effective for removing deuterons. For the large angle events, the
most important correlation was $E(t)-\theta(\text{heavy})$; here
``heavy'' means $Z=3$ particles detected in the gas detector volume.

\begin{figure}[htbp]
\includegraphics{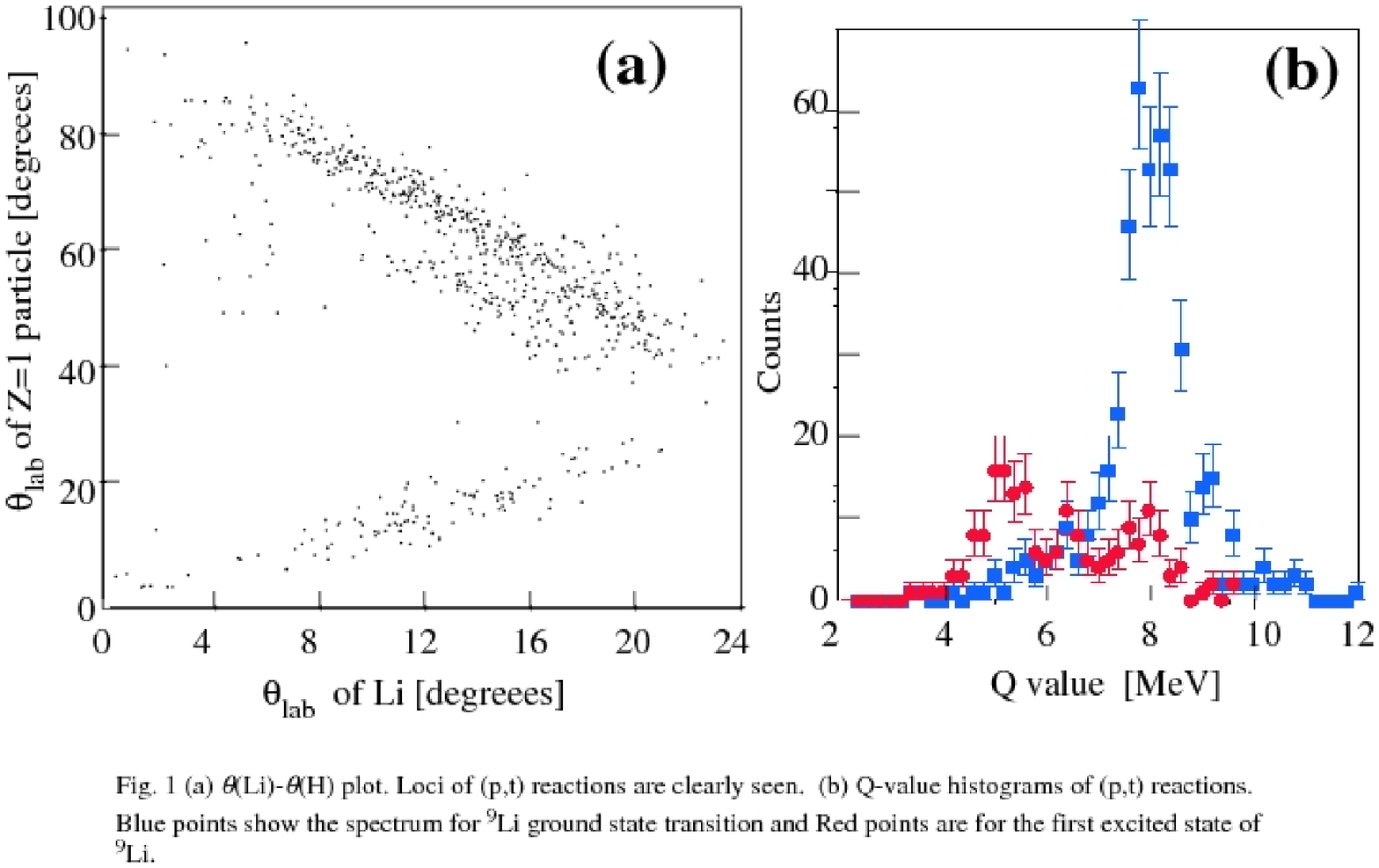}
\caption{\label{fig:fig1} (a) $\theta(\text{Li})-\theta(\text{H})$
plot. Loci of (p,t) reactions are clearly seen.  (b) $Q$-value
histograms of (p,t) reactions. Blue points show the spectrum for
\nuc{9}{Li} ground state transition and Red points are for the first
excited state of \nuc{9}{Li}.}
\end{figure}

Figure \ref{fig:fig1}(a) shows $\theta(\text{heavy})$ vs. $\theta(\text{light})$
scatter plot after those selections. Two clear kinematic loci are
seen. The reaction $Q$-value spectra calculated from those angles
are shown in the panel (b). The spectra show the transitions to the
ground state of \nuc{9}{Li} as well as the transition to the first
excited state \nuc{9}{Li} (2.69 MeV). Mixing of ground state
transition into the first excited state spectrum is seen in the plot
but, can easily be removed by the selection of the $Q$ values. The
tracking efficiency was determined by comparing the number of
identified particles (by the $\Delta{}E-E$ method) and the number of
tracks that hit an array detector at consistent position; this
comparison was undertaken separately for Li and triton ions.  The
geometrical efficiency was estimated by a Monte-Carlo simulation
that includes detector geometries and energy losses of the charged
particles in the gas.

The number of incident \nuc{11}{Li} ions was determined by counting
the incident ions both by the PPAC and signals from the first 7 cm
of the gas detector. The position, direction, and energy loss along
the track of an incident particle were used to select good incident
\nuc{11}{Li} ions. The uncertainty of the incident beam intensity is
$<1\%$. The largest uncertainty in the absolute value of the cross
section comes from the tracking efficiency and this is estimated to
be about $\pm10\%$.

\begin{figure}[htbp]
\includegraphics{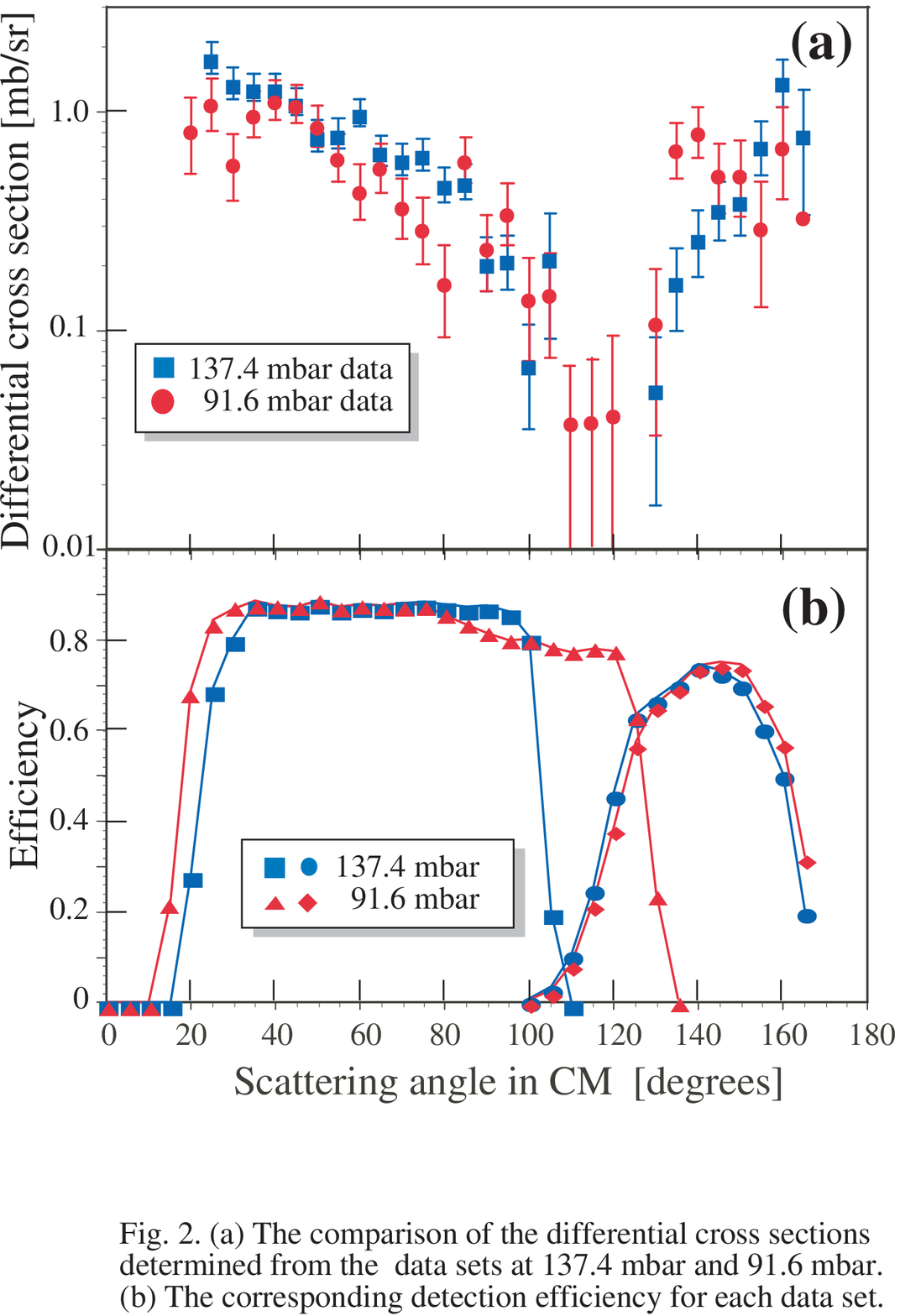}
\caption{\label{fig:fig2} (a) The comparison of the differential
cross sections determined from the data sets at 137.4 mbar and 91.6
mbar. (b) The corresponding detection efficiency for each data set.}
\end{figure}

Figure \ref{fig:fig2}(a) shows the determined differential cross sections from
the measurements with gas pressure equivalent to 137.4 mbar and 91.6
mbar at $0^\circ$C. The center of mass scattering angles were
calculated from the scattering angles of \nuc{9}{Li} and the triton.
The detection efficiencies are shown in the panel (b), as a function
of the center of mass angle. At the 137.4 mbar, the detection
efficiency drops to zero near $\theta_{cm}=110^\circ$; this is
because neither the \nuc{9}{Li} nor the triton ions reach the array
detector near $\theta_{cm}=110^\circ$. The efficiency of event
detection near cm=110бу was higher for the 91.6 mbar setting; under
this condition, either the \nuc{9}{Li} ion or the triton will hit
the array detector for all scattering angles.

\begin{figure}[htbp]
\includegraphics{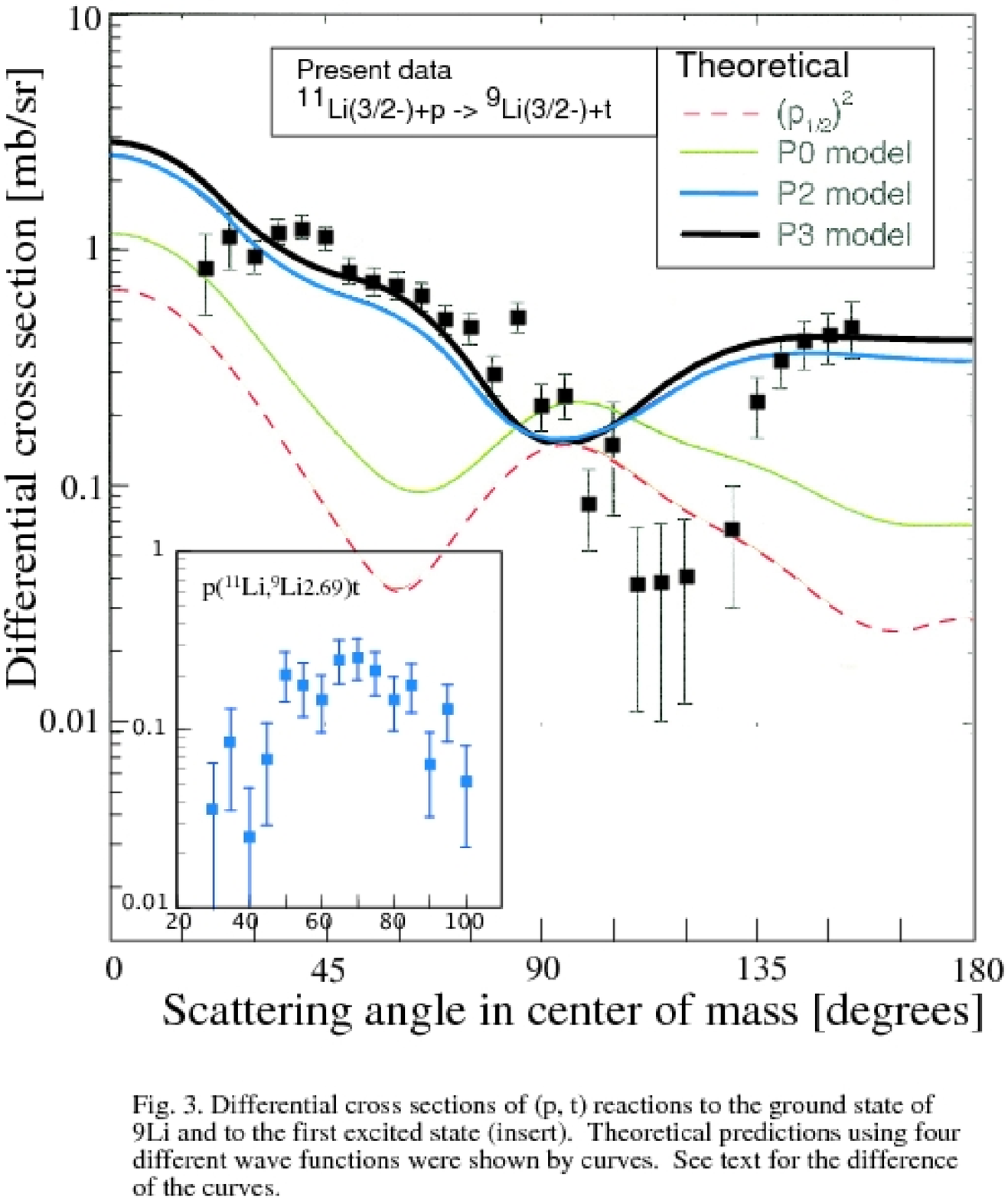}
\caption{\label{fig:fig3} Differential cross sections of (p,t)
reaction to the ground state of \nuc{9}{Li} and to the first excited
state (insert). Theoretical predictions using four different wave
functions were shown by curves. See text for the difference of the
wave functions.}
\end{figure}

For any particular
\nuc{11}{Li} reaction event, the incident energy depends on the
depth of the reaction point within the gas. In the present
experiment cross sections were averaged over \nuc{11}{Li} energies
from 2.8$A$ to 3.2$A$ MeV. The deduced differential cross sections
corresponding to the two different pressure settings were consistent
within experimental uncertainties. The averaged differential cross
sections for transitions to the \nuc{9}{Li} ground state are shown
in Fig. \ref{fig:fig3}, where the error bars on the figure include only
statistical errors. The overall uncertainty in the absolute cross
section values is about $\pm10\%$.

The transition to the first excited state (Ex=2.69 MeV) has been
observed, and cross sections are shown also in Fig. \ref{fig:fig3}. If this state
were populated by a direct transfer, it would indicate that a $1^+$
or $2^+$ halo component is present in the ground state of
\nuc{11}{Li}($\frac{3}{2}^-$) because the spin-parity of the
\nuc{9}{Li} first excited state is $\frac{1}{2}^-$. This is new
information that has not yet been observed in any of previous
investigations. Compound nucleus contribution should be small: at
present energy, the angular distribution of compound decay must be
essentially isotropic, and hence the deep minima observed in the
angular distributions of the ground state and the first excited
state exclude the strong contribution. However, before a final
conclusion can be made, detailed studies of coupled channels and
sequential transfer effects need to be undertaken.

Multistep transfer calculations to determine the differential cross
sections to the ground state of \nuc{9}{Li} have been made. For
these calculations several of the three-body models from
Ref.\cite{ref-11}, recalculated using the hyperspherical harmonic
expansions of Ref.\cite{ref-12}, with projection operators to remove
the $0\text{s}_{1/2}$ and $0\text{p}_{3/2}$ Pauli blocked states, have been used.
In particular, the P0, P2 and P3 models from \cite{ref-11},
which have percentage $(1\text{s}_{1/2})^2$ components of 3\%, 31\% and
45\%, respectively were used. The corresponding matter radii for
\nuc{11}{Li} are 3.05, 3.39 and 3.64 fm. For comparison, a simple
$(\text{p}_{1/2})^2$ model based on the P0 case, but with no n-n potential
to correlate the neutrons, was also investigated. All models here do
not include an excitation of \nuc{9}{Li} core.

The calculations reported here included the simultaneous transfer of
two neutrons from \nuc{11}{Li} to \nuc{9}{Li} in a one step process,
as well as coherently the two-step sequential transfers via
\nuc{10}{Li}. The simultaneous transfers used a triton wavefuction
calculated in the hyperspherical framework with the SSC(C)
nucleon-nucleon force \cite{ref-13}, and a three-body force to
obtain the correct triton binding energy. The sequential transfers
passed through both $\frac{1}{2}^+$ and $\frac{1}{2}^-$ neutron
states of \nuc{10}{Li}, with spectroscopic factors given by
respectively the s- and p-wave occupation probabilities for
\nuc{11}{Li} models of \cite{ref-11}. The spectroscopic amplitudes
for $\langle{}\text{d}|\text{t}\rangle$ and
$\langle\nuc{10}{Li}|\nuc{11}{Li}\rangle$ include a factor of $\sqrt{2}$ to
describe the doubled probability when either one of the two neutrons
can be transferred. S and P wave radial states were used with
effective binding energies of 1.0 and 0.10 MeV respectively; this
ensured a rms radii of $\sim6$ fm, which is the mean n-\nuc{9}{Li}
distance in the \nuc{11}{Li} models. The proton, deuteron and triton
channel optical potentials used are shown in Table \ref{table-1}. The
differential cross sections were obtained using the
FRESCO\cite{ref-xiv}.

\begin{table}
\caption{\label{table-1}Optical potential parameters used for the present calculations.}
\begin{tabular}{|r|r|r|r|r|r|r|r|r|r|r|}
\hline        & $V$ MeV & $r_V$ fm & $a_V$ fm & $W$ MeV & $W_D$ MeV & $r_W$ fm & $a_W$ fm & $V_{so}$ MeV & $r_{so}$ fm & $a_{so}$ fm \\
\hline p+\nuc{11}{Li}\cite{ref-xv} & 54.06 &  1.17 &  0.75 &  2.37 &  16.87 &  1.32 &  0.82 &     6.2 &   1.01 &   0.75 \\
\hline d+\nuc{10}{Li}\cite{ref-xvi} &  85.8 &  1.17 &  0.76 & 1.117 & 11.863 & 1.325 & 0.731 &       0 &        &        \\
\hline  t+\nuc{9}{Li}\cite{ref-xvii} &  1.42 &  1.16 &  0.78 &  28.2 &      0 &  1.88 &  0.61 &       0 &        &        \\
\hline
\end{tabular}
\end{table}

Curves in Fig. \ref{fig:fig3} show the results of the calculations. The wave
function $(\text{p}_{1/2})^2$ with no n-n correlation gives very small
cross sections that are far from the measured values. Also the P0
wave function, with n-n correlation but with a small $(\text{s}_{1/2})^2$
mixing amplitude, gives too small cross sections. The results of the
P2 and P3 wave functions fit the forward angle data reasonably
well but the fitting near the minimum of the cross section is
unsatisfactory. The results may be sensitive to the choice of the
optical potentials as well as the selection of the intermediate
states of two-step processes. Detailed analysis of such effects
should be a subject of future studies.

In summary, we have measured for the first time the differential
cross sections for two-halo neutron transfer reactions of the most
pronounced halo nucleus \nuc{11}{Li}.  Transitions were observed to
the ground and first excited state of \nuc{9}{Li}. Multistep
transfer calculations were applied with different wave functions of
\nuc{11}{Li}. It is seen that wave functions with strong mixing of p
and s neutrons which includes three-body correlations, provides the
best fit to the data for the magnitude of the reaction cross
section. However the fitting to the angular shape is less
satisfactory. The population of the first excited state of
\nuc{9}{Li} suggests a $1^+$ or $2^+$ configuration of the halo
neutrons. This shows that a two-nucleon transfer reaction as studied
here may give a new insight in the halo structure of \nuc{11}{Li}.
Further studies clearly are necessary to understand the observed
cross sections as well as the correlation between the two-halo
neutrons.

One of the authors, IT, acknowledges the support of TRIUMF
throughout his stay at TRIUMF. The experiment is supported by GANIL
and technical help from J. F. Libin, P. Gangnant, C. Spitaels, L.
Olivier, and G. Lebertre are gratefully acknowledged. This work was
supported by the NSERC of Canada through TRIUMF and Saint Mary's
University. Part of this work was performed under the auspices of
the U.S. Department of Energy by Lawrence Livermore National
Laboratory under Contract DE-AC52-07NA27344. This experiment was the
first experiment at the new ISAC-2 facility.  The authors gratefully
acknowledge R. Laxdal, M. Marchetto, M. Dombsky and all other staff
members for their excellent effort for setting up the beam line and
delivering the high-quality \nuc{11}{Li} beam.


\begin{thebibliography}{99}
\bibitem{ref-01} I. Tanihata et al., Phys. Letters B \textbf{287} (1992) 307.
\bibitem{ref-02} M. Zinser et al., Nucl. Phys. A \textbf{619} (1997) 151.
\bibitem{ref-03} S. Shimoura et al., Phys. Letters B \textbf{348} (1995) 29.
\bibitem{ref-04} T. Nakamura et al., Phys. Rev. Letters \textbf{96} (2006) 252502.
\bibitem{ref-05} R. Sanchez et al., Phys. Rev. Letters \textbf{96} (2006) 033002.
\bibitem{ref-06} I. Tanihata et al., Phy. Rev. Letters \textbf{55} (1985) 2676. A.Ozawa, T. Suzuki, and I. Tanihata., Nucl. Phys. A \textbf{693} (2001) 32.
\bibitem{ref-vii} L. -B Wang et al., Phys. Rev. Letters \textbf{93} (2004) 142501.
\bibitem{ref-viii} M. V. Zhukov et al., Phys. Rep. \textbf{231} (1993) 151.
\bibitem{ref-09} Y. T. V. Oganessian et al., Phys. Rev. Letters \textbf{82} (1999) 4996.
\bibitem{ref-10} C. E. Demonchy et al., Nuc. Instr \& Methods A \textbf{573} (2007) 145.
\bibitem{ref-11} I.J. Thompson and M.V. Zhukov, Phys. Rev. C \textbf{49} (1994) 1904.
\bibitem{ref-12} I.J. Thompson et al., Phys. Rev. C \textbf{61}, 24318 (2000).
\bibitem{ref-13} R. de Tourreil and D.W.L. Sprung, Nucl. Phys. A \textbf{242} (1975) 445.
\bibitem{ref-xiv} I. J. Thompson, Computer Physics Report \textbf{7} (1988) 167.
\bibitem{ref-xv}  F. D. Beccehetti and G. W. Greenless, Phys. Rev. \textbf{182} (1969) 1190.
\bibitem{ref-xvi} W. W. Daehnick, J. D. Childs,and Z. Vrcelj  Phys. Rev. C \textbf{21} (1980) 2253.
\bibitem{ref-xvii} F.D. Becchetti and G.W. Greenlees, Annual Report, J. H. Williams Laboratory, University of Minnesota (1969).
\end{thebibliography}
\end{document}